\documentclass[aps,twocolumn,showpacs,prb]{revtex4}

\usepackage{dcolumn}
\usepackage{amsmath}
\usepackage{graphicx}

\begin{document}

\title{Planar approximation for spin transfer systems  \\
 with application to tilted polarizer devices}

\author{Ya.\ B. Bazaliy}
 \affiliation{Department of Physics and Astronomy, University of South Carolina, Columbia, SC,}
 \affiliation{Institute of Magnetism, National Academy of Science, Ukraine.}

\date{\today}

\begin{abstract}
Planar spin-transfer devices with dominating easy-plane anisotropy
can be described by an effective one-dimensional equation for the
in-plane angle. Such a description provides an intuitive qualitative
understanding of the magnetic dynamics. We give a detailed
derivation of the effective planar equation and use it to describe
magnetic switching in devices with tilted polarizer.
\end{abstract}


\maketitle

\section{Introduction}

The spin-transfer effect is a non-equilibrium interaction that
arises when a current of electrons flows through a non-collinear
magnetic texture \cite{berger,slon96,bjz1997}. Spin transfer torque
can lead to current induced magnetic switching in multilayer devices
or domain wall motion in devices with continuous change of
magnetization. Both phenomena serve as an underlying mechanism for a
number of suggested memory and logic applications.

Magnetic dynamics in spin-transfer devices can be described by the
Landau-Lifshitz-Gilbert (LLG) equation. Analytic solutions of LLG
can be easily found in the simplest case of easy axis magnetic
anisotropy. However, when the form of anisotropy energy becomes more
complicated the investigations of the stability of static equilibria
become much more involved. A study of the precession cycles is even
more complicated and often makes it necessary to resort to numeric
simulations. Due to the complexity of the LLG equation it is always
interesting to consider cases where some simplifying approximations
can be made.

\begin{figure}[b]
    \resizebox{.45\textwidth}{!}{\includegraphics{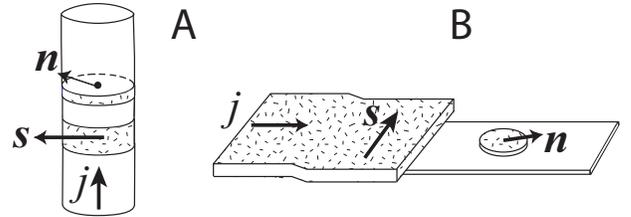}}
\caption{Planar spin-transfer devices. Hashed parts of the devices
are ferromagnetic, clear parts are made from a non-magnetic metal.}
 \label{fig:devices}
\end{figure}

In many devices the easy plane anisotropy energy is much larger than
the other anisotropy energies, and the system is in the planar
spintronic device regime \cite{bauer-planar-review}
(Fig.~\ref{fig:devices}). The limit of dominating easy plane energy
is characterized by a simplification of the dynamic equations
\cite{weinan-e}, which comes not from the high symmetry of the
problem, but from the existence of a small parameter: the ratio of
the energy modulation within the plane to the easy plane energy. The
deviation of the magnetization from the plane becomes small, making
the motion effectively one dimensional. As a result, an effective
description in terms of just one azimuthal angle becomes possible.

Our publications~\cite{boj:2008, bazaliy:2007:APL,
bazaliy:2007:PRB,bazaliy:2009:SPIE} introduced the planar
approximation in the presence of spin-transfer torques and listed a
number of results highlighting the practical use of the method. In
this paper we give a detailed derivation the effective planar
equation for a macrospin free layer in the presence of spin transfer
torques (Sec.~\ref{sec:effective_equations}). We then show in
Sec.~\ref{sec:tilted_polarizer_device} how this equation can be
applied to a system with the tilted polarizer and obtain a
qualitative picture of the device dynamics.

\section{Magnetic dynamics of the free layer}

We consider a conventional spin-transfer device consisting of a a
magnetic polarizer (fixed layer) and a small magnet (free layer)
with electric current flowing from one to another
(Fig.~\ref{fig:devices}). The free layer is influenced by the spin
transfer torque, while the polarizer is too large to feel it. It is
assumed that due to the large exchange stiffness the free layers can
be described by a macrospin model. Its magnetic dynamics is
described by the Landau-Lifshitz-Gilbert equation with the spin
transfer torque term.

The state of the free layer is characterized by just one vector, its
total magnetic moment ${\bf M} = M {\bf n}$ which has a constant
absolute value $M$ and a direction given by a unit vector ${\bf
n}(t)$. The LLG equation \cite{slon96,bjz2004} reads:
\begin{equation}
 \label{eq:vector_LLG}
{\dot {\bf n}} = \left[ - \frac{\delta \varepsilon}{\delta {\bf n}}
\times {\bf n} \right] + u({\bf n}) [{\bf n} \times [{\bf s} \times
{\bf n}]] + \alpha [{\bf n} \times \dot {\bf n}] \ .
\end{equation}
Here the re-scaled energy $\varepsilon = (\gamma/M) E$ has the
dimensions of frequency and is expressed through the total magnetic
energy $E({\bf n})$ of the free layer; $\gamma$ is the gyromagnetic
ratio, and $\alpha$ is the Gilbert damping constant. The second term
on the right is the spin transfer torque. Unit vector ${\bf s}$
points along the direction of the polarizer moment, and the spin
transfer strength $u({\bf n})$ is proportional to the electric
current $I$ \cite{bjz2004}. In general, spin transfer strength is a
function of the angle between the polarizer and the free layer
$u({\bf n}) = g[({\bf n}\cdot{\bf s})] \ I$. The spin current
efficiency factor $g[({\bf n}\cdot{\bf s})]$ is a material and
device specific function.

In polar angles $(\theta, \phi)$ equation (\ref{eq:vector_LLG})
reads
\begin{eqnarray}
 \nonumber
\dot\theta + \alpha \dot\phi \sin\theta &=&
    - \frac{1}{\sin\theta} \frac{\partial \varepsilon}{\partial\phi}
    + u({\bf n}) ({\bf s} \cdot {{\bf e}_{\theta}}) \ ,
 \\
 \label{eq:polar_angles_LLG}
\dot\phi \sin\theta - \alpha \dot\theta &=&
    \frac{\partial \varepsilon}{\partial\theta}
    +  u({\bf n}) ({\bf s} \cdot {\bf e}_{\phi}) \ ,
\end{eqnarray}
where the unit tangent vectors ${\bf e}_{\theta}$ and ${\bf
e}_{\phi}$ are defined in the
Appendix~\ref{appendix:vector_definitions}.

We assume that the easy plane is defined by $\theta = \pi/2$, and
the rescaled magnetic energy has the form
$$
\varepsilon = \frac{\omega_p}{2}\cos^2\theta +
\varepsilon_r(\theta,\phi) \ .
$$
The first term represents the easy plane anisotropy, with the planar
frequency $\omega_p$ related to the easy plane constant $K_p$ as
$\omega_p = \gamma K_p/M$. The remainder $\varepsilon_r$ is the
``residual'' energy. The planar limit is characterized by $\omega_p
\to \infty$. Large easy plane constant forces the energy minima to
be very close to the easy plane and the low energy solutions of LLG
to have the property $\theta(t) = \pi/2 + \delta\theta$ with
$\delta\theta \to 0$. Equations~(\ref{eq:polar_angles_LLG}) can then
be expanded in small parameters
$$
\frac{|\varepsilon_r|}{\omega_p} \ll 1 \quad {\rm and}
 \quad \frac{|u({\bf n})|}{\omega_p} \ll 1 \ .
$$
By truncating this expansion one obtains the effective planar
approximation.

\section{Derivation of the effective planar equation}
\label{sec:effective_equations}

Explicitly separating the large easy plane terms, we rewrite
equation (\ref{eq:polar_angles_LLG}) as
\begin{eqnarray*}
\dot\theta + \alpha \dot\phi \sin\theta &=& f_{\theta} + u_{\theta}
 \\
\dot\phi \sin\theta - \alpha \dot\theta &=& -\omega_p
\cos\theta\sin\theta + f_{\phi} + u_{\phi}
\end{eqnarray*}
where the residual anisotropy is responsible for the terms
\begin{eqnarray}
 \nonumber
 f_{\theta}(\theta,\phi) &=& -\frac{1}{\sin\theta}
 \frac{\partial \varepsilon_r}{\partial\phi} \ ,
 \\ \label{eq:definitions_f_theta_phi}
 f_{\phi}(\theta,\phi) &=& \frac{\partial \varepsilon_r}{\partial\theta} \ ,
\end{eqnarray}
and the spin transfer torque produces the terms
\begin{eqnarray}
 \nonumber
 u_{\theta}(\theta,\phi) &=& u({\bf n}) ({\bf s} \cdot {\bf e}_{\theta}) \
 ,
\\ \label{eq:definitions_u_theta_phi}
 u_{\phi}(\theta,\phi) &=& u({\bf n}) ({\bf s} \cdot {\bf e}_{\phi}) \ .
\end{eqnarray}
We also introduce a notation $F_{\theta,\phi} = f_{\theta,\phi} +
u_{\theta,\phi}$, and re-write the LLG system as
\begin{eqnarray}
 \nonumber
\dot\theta &=& \frac { F_{\theta} - \alpha(-\omega_p
\cos\theta\sin\theta + F_{\phi}) } {1+\alpha^2}
\\
 \label{eq:dotthetadotphi}
 &&
\\
 \nonumber
 \dot\phi &=& \frac { -\omega_p \cos\theta\sin\theta
 + F_{\phi} + \alpha F_{\theta}} {(1 + \alpha^2)\sin\theta}
\end{eqnarray}

Next, we make the approximations. In the $\omega_p \to \infty$ limit
the solution is expected to have a property $\theta(t) = \pi/2 +
\delta\theta(t)$ with $\delta\theta \to 0$. Expanding all quantities
on the r.h.s. of (\ref{eq:dotthetadotphi}) in small $\delta\theta$
up to the first order we get
\begin{eqnarray}
 \label{eq:LLG_first_order_in_theta_A}
 \delta{\dot\theta} &=& \frac { F^{0}_{\theta} + F^{1}_{\theta} \delta\theta -
 \alpha(\omega_p \delta\theta + F^{0}_{\phi} + F^{1}_{\phi}\delta\theta) } {1+\alpha^2} \ ,
\\
 \label{eq:LLG_first_order_in_theta_B}
 \dot\phi &=& \frac { \omega_p \delta\theta + F^{0}_{\phi} + F^{1}_{\phi}\delta\theta
 + \alpha (F^{0}_{\theta} + F^{1}_{\theta} \delta\theta) } {1 + \alpha^2}
 \ ,
\end{eqnarray}
where we have used the notation
\begin{eqnarray*}
F^{0} &=& F \left( \frac{\pi}{2},\phi \right) \ ,
 \\
F^{1} &=& \frac{\partial F}{\partial\theta}
 \left( \frac{\pi}{2},\phi \right)  \ .
\end{eqnarray*}
In the approximation
(\ref{eq:LLG_first_order_in_theta_A},\ref{eq:LLG_first_order_in_theta_B})
equations are linear with respect to the unknown function
$\theta(t)$, but still fully non-linear with respect to $\phi(t)$.

Eq.~(\ref{eq:LLG_first_order_in_theta_B}) can be viewed as an
equation on $\delta\theta$ which gives
\begin{equation}\label{eq:slave_condition}
\delta\theta = \frac{
    (1+\alpha^2)\dot\phi - F^{0}_{\phi} - \alpha F^{0}_{\theta}
    }{
    \omega_p + F^{1}_{\phi} + \alpha F^{1}_{\theta}
    } = q(\phi,\dot\phi)
\end{equation}
so that the out-of-plane deviation becomes a ``slave'' of the
in-plane motion.\cite{weinan-e} The presence of the large $\omega_p$
in the denominator ensures the smallness of $\delta\theta$.
Substituting the resulting expression $\delta\theta =
q(\phi,\dot\phi)$ back into equation
(\ref{eq:LLG_first_order_in_theta_A}) one obtains a second order
differential equation for a single unknown function $\phi(t)$
$$
  \frac{\partial q}{\partial \dot\phi} \ \ddot\phi
    +
    \frac{\partial q}{\partial \phi} \ \dot\phi
  =
    \frac{F^0_{\theta} - \alpha F^0_{\phi}}{1+\alpha^2} +
    \left(
    \frac{F^1_{\theta} - \alpha\omega_p - \alpha F^1_{\phi}}{1+\alpha^2}
    \right) q \ .
$$
By denoting $\Omega(\phi) = \omega_p + F^1_{\phi} + \alpha
F^1_{\theta}$ and simplifying the terms we get
\begin{eqnarray}
    \nonumber
  \frac{1+\alpha^2}{\Omega} \ddot\phi
  &+&
    \left(
      \alpha + \frac{\partial q}{\partial \phi} - (1+\alpha^2)
      \frac{F^1_{\theta}}{\Omega}
    \right) \dot\phi =
    \\ \label{eq:ddot_phi_exact}
& = & F^0_{\theta} -
    \frac{F^1_{\theta} (F^0_{\phi} + \alpha F^0_{\theta})}{\Omega}
\end{eqnarray}
In this form the equation is still too complicated to be useful but
since it was obtained from the approximations
(\ref{eq:LLG_first_order_in_theta_A},\ref{eq:LLG_first_order_in_theta_B})
we are allowed to drop the terms that are smaller or equal to the
already neglected ones. Those were the $F'' \delta\theta^2$ terms in
the expansion of $F$, and the third order terms $\omega_p
\delta\theta^3$ in the easy plane energy expansion. While formally
the latter are of higher order in the $\delta\theta$ expansion, the
large coefficient $\omega_p$ can causes them to have the same order
of magnitude.

To compare the orders of magnitude of the terms consistently, we
need to know the order of magnitude of $\delta\theta$. At the
present stage we know that $\delta\theta$ is small but
its exact order of magnitude is not known because we do not have an
estimate for the $\dot\phi$ term in the numerator of
Eq.~(\ref{eq:slave_condition}).

\subsection{Simple residual energy in the absence of spin torque}
\label{sec:simple_E_zero_current}

To estimate $\delta\theta$, let us first consider the problem in the
absence of spin transfer ($F = f$),\cite{weinan-e} assuming a simple
form of residual energy $\varepsilon_r = \varepsilon_r(\phi)$. In
this case we find $f^0_{\phi} = f^1_{\phi} = f^1_{\theta} = 0$ and
get
\begin{equation}\label{eq:delta_theta_no_ST}
q(\phi,\dot\phi) = \frac{(1+\alpha^2)\dot\phi - \alpha f^0_{\theta}
}{\omega_p} \ .
\end{equation}
Equation (\ref{eq:ddot_phi_exact}) takes a form
$$
\frac{1+\alpha^2}{\omega_p}\ddot\phi + \alpha\left( 1 -
 \frac{1}{\omega_p} \frac{\partial f^0_{\theta}}{\partial\phi} \right)\dot\phi
 = f^0_{\theta} \ .
$$
Assuming that the residual energy $\varepsilon_r(\phi)$ does not
have any special points of fast change we can estimate $F^0_{\theta}
\sim \varepsilon \ll \omega_p$. Then
$$
1 - \frac{1}{\omega_p} \frac{\partial f^0_{\theta}}{\partial\phi}
\approx 1
$$
and we can approximate the equation by
$$
\frac{1+\alpha^2}{\omega_p}\ddot\phi + \alpha\dot\phi
 = f^0_{\theta} = -\frac{\partial\varepsilon_r}{\partial\phi} \ .
$$
The equation above has the form of the Newton's equation for a
particle of mass $(1+\alpha^2)/\omega_p$ moving in a one-dimensional
potential $\varepsilon_r(\phi)$, subject to a viscous friction force
with a friction coefficient $\alpha$. Our goal is to estimate the
value of $\dot\phi$, i.e., the speed of the ``effective particle''.

The particle's characteristic speed depends on the total energy and
on the relative strength of the friction forces. We will assume that
the energy is of the order of $\varepsilon_r$ (this is the
mathematical equivalent of our original assumption about the
low-energy dynamics of the moment). Furthermore, in the present
paper we will concentrate on the case of $\alpha \to 0$ that
corresponds to an almost frictionless motion of the particle. Then
one can use the approximate energy conservation and write
$(1/2\omega_p) \dot\phi^2 = \varepsilon_r$ for the maximum speed.
This gives
$$
\dot\phi \sim \sqrt{\varepsilon_r \omega_p}
$$
Using similar arguments one can estimate the maximum acceleration as
$$
\ddot\phi \sim \varepsilon_r \omega_p
$$
Note that the viscous friction can be approximately neglected when
$\alpha\dot\phi \ll F^0_{\theta} \sim \varepsilon_r$, i.e., the
Gilbert damping constant $\alpha$ has to be not just small compared
to unity but satisfy a more stringent inequality
\begin{equation}\label{eq:small_alpha_condition}
\alpha \ll \sqrt{\frac{\varepsilon_r}{\omega_p}} \ll 1 \ .
\end{equation}
We see now that $\dot\phi$ is the largest term in the denominator of
(\ref{eq:delta_theta_no_ST}) and as a result obtain an estimate
\begin{equation}\label{eq:order_delta_theta_no_ST}
\delta\theta  \sim \sqrt{\frac{\varepsilon_r}{\omega_p}} \ .
\end{equation}

\subsection{Arbitrary residual energy in the absence of spin
torque}\label{arbitrary_E_zero_current}

Let us return to the approximation
(\ref{eq:LLG_first_order_in_theta_A},\ref{eq:LLG_first_order_in_theta_B})
with the general form of the residual energy $\varepsilon_r =
\varepsilon_r(\theta,\phi)$ and zero current, $u = 0$, $F = f$. It
is now possible to use the {\em a posteriori} estimate
(\ref{eq:order_delta_theta_no_ST}) for $\delta\theta$ to consider
the orders of magnitude of the terms on the right hand sides of the
equations. We start the discussion from the ``slave''
equation~(\ref{eq:LLG_first_order_in_theta_B}). Here
\begin{eqnarray*}
\omega_p \delta\theta & \sim & \sqrt{\varepsilon_r \omega_p}
 \\
f^0 & \sim & \varepsilon_r
 \\
f^1 \delta\theta & \sim & \varepsilon_r
\sqrt{\frac{\varepsilon_r}{\omega_p}}
 \\
\alpha f^0 & \ll & \varepsilon_r
\sqrt{\frac{\varepsilon_r}{\omega_p}}
 \\
\alpha f^1 \delta\theta & \ll & \varepsilon_r
\frac{\varepsilon_r}{\omega_p}
\end{eqnarray*}
As we see, the orders of magnitude of the terms form a series
\begin{equation}\label{eq:term_order_series}
\ldots
 \quad \varepsilon_r \frac{\varepsilon_r}{\omega_p}
 , \quad \varepsilon_r \sqrt{\frac{\varepsilon_r}{\omega_p}}
 , \quad \varepsilon_r
 , \quad \sqrt{\varepsilon_r \omega_p}
 \quad \ldots
\end{equation}
with the general term given by $\varepsilon_r
(\varepsilon_r/\omega_p)^{n/2}$.

The terms neglected in transition from system
(\ref{eq:dotthetadotphi}) to (\ref{eq:LLG_first_order_in_theta_B})
were
\begin{eqnarray*}
\omega_p \delta\theta^3 & \sim & \varepsilon_r
\sqrt{\frac{\varepsilon_r}{\omega_p}} \ ,
 \\
\frac{\partial^2 f}{\partial\theta^2} \delta\theta^2 & \sim &
\varepsilon_r \frac{\varepsilon_r}{\omega_p} \ .
\end{eqnarray*}
This means that in (\ref{eq:LLG_first_order_in_theta_B}) one should
only keep the terms of the order $\varepsilon_r$ and higher. Lower
order terms would be comparable to some of the discarded terms.
Using this argument we discard $f^1 \delta\theta$, $\alpha f^0$ and
$\alpha f^1 \delta\theta$. The $1/(1+\alpha^2)$ factor in
(\ref{eq:LLG_first_order_in_theta_B}) can be expanded using
Eq.~(\ref{eq:small_alpha_condition})
$$
\frac{1}{1+\alpha^2} = 1 + \delta, \quad
 \delta \sim \alpha^2 \ll \frac{\varepsilon_r}{\omega_p} \ .
$$
This inequality shows that $1/(1+\alpha^2)$ can can be approximated
by unity in Eq.~(\ref{eq:LLG_first_order_in_theta_B}) without
changing the accuracy. After those simplifications equation
(\ref{eq:slave_condition}) takes the form
$$
q(\phi,\dot\phi) = \frac{\dot\phi - f^{0}_{\phi}}{\omega_p} \ ,
$$

As for the equation~(\ref{eq:LLG_first_order_in_theta_A}), the terms
discarded in going from (\ref{eq:dotthetadotphi}) to
(\ref{eq:LLG_first_order_in_theta_A}) were
\begin{eqnarray*}
\alpha \omega_p \delta\theta^3 & \ll & \varepsilon_r
\frac{\varepsilon_r}{\omega_p}
 \\
\frac{\partial^2 f}{\partial\theta^2} \delta\theta^2 & \sim &
\varepsilon_r \frac{\varepsilon_r}{\omega_p}
\end{eqnarray*}
and therefore we have to keep the terms of the orders $\varepsilon_r
\sqrt{\varepsilon_r / \omega_p}$ and higher. Thus the $f^1
\delta\theta$ and $\alpha f^0$ terms should be kept in
(\ref{eq:LLG_first_order_in_theta_A}) but the $\alpha F^1
\delta\theta$ terms should be discarded. One can also conclude that
it is safe to replace the $1/(1+\alpha^2)$ factor by unity.
Equation~(\ref{eq:LLG_first_order_in_theta_A}) is now replaced by
$$
 \delta{\dot\theta} = f^{0}_{\theta} + f^{1}_{\theta} \delta\theta -
 \alpha(\omega_p \delta\theta + f^{0}_{\phi})  \ ,
$$
where one should use
$$
f^{1}_{\theta} \delta\theta = f^{1}_{\theta} \frac{\dot\phi -
f^{0}_{\phi}}{\omega_p} \approx  \frac{f^{1}_{\theta}
\dot\phi}{\omega_p} \ ,
$$
since the term $f^{1}_{\theta} f^{0}_{\phi}/\omega_p \sim
\varepsilon_r^2/\omega_p$ is of the same order as the already
discarded terms.

Re-deriving Eq.~({\ref{eq:ddot_phi_exact}) in this approximation one
gets
$$
\frac{\ddot\phi}{\omega_p} + \left(\alpha - \frac{1}{\omega_p}
 \left[ \frac{\partial f^0_{\phi}}{\partial\phi} + f^1_{\theta}
 \right]\right)\dot\phi =
 f^0_{\theta} = -\frac{\partial\varepsilon_r(\pi/2,\phi)}{\partial\phi}
$$
This is the equation for the effective particle as discussed in the
previous section, except that the viscous friction coefficient seems
to acquire a correction. While the correction is of the order
$\varepsilon_r/\omega_p$, it is added to a small number $\alpha$ and
thus can potentially have a significant effect of changing the sign
of the dissipation term. However, one finds
\begin{equation}\label{eq:identity_on_f}
\frac{\partial f^0_{\phi}}{\partial\phi} + f^1_{\theta} =
 \frac{\partial}{\partial\phi}\frac{\partial \varepsilon_r}{\partial\theta}
 +
 \frac{\partial}{\partial\theta}
 \left(-\frac{\partial \varepsilon_r}{\partial\phi}\right) = 0 \ ,
\end{equation}
so the correction actually vanishes. We come back to the effective
equation
$$
\frac{\ddot\phi}{\omega_p} + \alpha \dot\phi =
-\frac{\partial\varepsilon_r(\pi/2,\phi)}{\partial\phi} \ ,
$$
which corresponds to the most natural generalization of the equation
derived in the previous section for a special form of the residual
energy $\varepsilon_r = \varepsilon_r(\phi)$. The positive effective
friction coefficient ensures that the effective particle always
stops in the point of energy minimum, as expected for a closed
system with dissipation described by the LLG equation without spin
torques.

\subsection{Effective equation in the presence of spin torque}
\label{sec:arbitrary_E_with_current}

Finally, we proceed to the derivation of the effective equation in
the presence of the spin torque. Consider approximations
(\ref{eq:LLG_first_order_in_theta_A},\ref{eq:LLG_first_order_in_theta_B})
with $\varepsilon_r = \varepsilon_r(\theta,\phi)$ and $u \neq 0$.

The order of magnitude of the extra terms produced by the current
will depend on the value of $u$. One condition that certainly has to
be satisfied is the smallness of the spin torque compared to the
anisotropy torques produced by the easy plane contribution. The
latter are responsible for the terms of the order
$\sqrt{\varepsilon_r \omega_p}$ in
Eqs~(\ref{eq:LLG_first_order_in_theta_A},\ref{eq:LLG_first_order_in_theta_B}).
Thus it seems that  $u$ should not exceed $\varepsilon_r$, which is
the largest term before $\sqrt{\varepsilon_r \omega_p}$ in the
series (\ref{eq:term_order_series}). Such a conclusion is correct
for a general situation. We will, however, see below that in some
special cases the current can be increased up to $u \sim
\sqrt{\varepsilon_r \omega_p}$ without violating the dominance of
the easy plane anisotropy torque.

To include those cases we assume $u \lesssim \sqrt{\varepsilon_r
\omega_p}$ and revisit
Eqs.~(\ref{eq:LLG_first_order_in_theta_A},\ref{eq:LLG_first_order_in_theta_B})
discarding the terms smaller than $\varepsilon_r
\sqrt{\varepsilon_r/\omega_p}$ in
Eq.~(\ref{eq:LLG_first_order_in_theta_A}), and smaller than
$\varepsilon_r$ in Eq.~(\ref{eq:LLG_first_order_in_theta_B}).

Eq.~(\ref{eq:LLG_first_order_in_theta_B}) which acquires the form
$$
\dot\phi = (\omega_p + u_{\phi}^1)\delta\theta
 + f^0_{\phi} + u^0_{\phi} \ .
$$
(as in the previous section, one can prove that the factor
$1/(1+\alpha^2)$ can be approximated by unity without loss of
accuracy). By solving for $\delta\theta$ and expanding the
denominator up to the same accuracy we find the form of the slave
condition (\ref{eq:slave_condition})
\begin{equation}\label{eq:g_with u}
\delta\theta =
 \left( 1 - \frac{u^1_{\phi}}{\omega_p} \right)\frac{\dot\phi}{\omega_p}
 - \frac{f^0_{\phi} + u^0_{\phi}}{\omega_p}
 + \frac{u^1_{\phi} u^0_{\phi}}{\omega_p^2}
\end{equation}
Differentiating both sides one gets
\begin{eqnarray}
 \nonumber
\delta\dot\theta &=&
 \left( 1 - \frac{u^1_{\phi}}{\omega_p} \right)
 \frac{\ddot\phi}{\omega_p}
 -  \frac{\partial u^1_{\phi}}{\partial\phi}
 \frac{\dot\phi^2}{\omega_p^2} -
\\ \label{eq:delta_dot_theta_with u}
 &&- \left(
   \frac{\partial f^0_{\phi}}{\partial\phi} + \frac{\partial u^0_{\phi}}{\partial\phi}
 - \frac{1}{\omega_p} \frac{\partial \big[u^1_{\phi} u^0_{\phi} \big]}{\partial
 \phi}
 \right)\frac{\dot\phi}{\omega_p}
\end{eqnarray}
Returning to Eq.~(\ref{eq:LLG_first_order_in_theta_A}) we find that
with the declared accuracy it can be rewritten as
$$
\delta\dot\theta = f^0_{\theta} + u^0_{\theta} - \alpha u^0_{\phi} +
 \left(f^1_{\theta} + u^1_{\theta} - \alpha\omega_p \right)
 \delta\theta \ .
$$
Substituting $\delta\theta$ from (\ref{eq:g_with u}) and discarding
any terms that are smaller than $\varepsilon_r
\sqrt{\varepsilon_r/\omega_p}$, we get
\begin{eqnarray*}
\delta\dot\theta &=& f^0_{\theta} + u^0_{\theta} -
 \left( \alpha - \frac{f^1_{\theta} + u^1_{\theta}}{\omega_p}
 + \frac{u^1_{\theta} u^1_{\phi}}{\omega_p^2}  \right)\dot\phi -
 \\
&& - \frac{f^1_{\theta} u^0_{\phi} + u^1_{\theta} f^0_{\phi} +
u^1_{\theta} u^0_{\phi}}{\omega_p} +
 \frac{u^1_{\theta} u^1_{\phi} u^0_{\phi}}{\omega_p^2}
\end{eqnarray*}
The last step is to use Eq.~(\ref{eq:delta_dot_theta_with u}) to
express $\delta\dot\theta$ on the left hand side. This gives the
form of the effective equation (\ref{eq:ddot_phi_exact}) without the
terms below our accuracy
\begin{eqnarray*}
&& \left( 1 - \frac{u^1_{\phi}}{\omega_p} \right)
   \frac{\ddot\phi}{\omega_p} +
 \\
&& + \left(
    \alpha - \frac{1}{\omega_p}\left[
         \frac{\partial f^0_{\phi}}{\partial\phi} + f^1_{\theta}
         + \frac{\partial u^0_{\phi}}{\partial\phi} + u^1_{\theta}
         \right]  -
         \right.
 \\
&&       \left.
 \quad\quad \ - \frac{1}{\omega_p^2} \left[
  \frac{\partial \big[u^1_{\phi} u^0_{\phi} \big]}{\partial \phi}
    - u^1_{\theta} u^1_{\phi}
        \right]
 \right)\dot\phi
 -  \frac{\partial u^1_{\phi}}{\partial\phi}
 \frac{\dot\phi^2}{\omega_p^2} =
 \\
&& = f^0_{\theta} + u^0_{\theta}
  - \frac{f^1_{\theta} u^0_{\phi} + u^1_{\theta} f^0_{\phi} +
  u^1_{\theta} u^0_{\phi}}{\omega_p} +
 \frac{u^1_{\theta} u^1_{\phi} u^0_{\phi}}{\omega_p^2} \ ,
\end{eqnarray*}
where identity (\ref{eq:identity_on_f}) can be used to simplify the
bracketed expression on the second line.

We now cast the effective planar equation in its final form
\begin{equation}\label{eq:effective_planar_equation}
m \ddot\phi + \alpha_{eff}\dot\phi
 -  \frac{(u^1_{\phi})'}{\omega_p^2}\dot\phi^2
 = -(\varepsilon_{eff})' \ .
\end{equation}
Here primes denote differentiation with respect to $\phi$, and the
parameters are given by
\begin{eqnarray}
 \nonumber
 m &=& \frac{1}{\omega_p} \left( 1 - \frac{u^1_{\phi}}{\omega_p}
 \right) \ ,
\\ \label{eq:effective_parameters}
 \alpha_{eff} &=& \alpha -
 \frac{ (u^0_{\phi})' + u^1_{\theta} }{\omega_p}
  + \frac{\big( u^1_{\phi} u^0_{\phi} \big)'
  + u^1_{\phi} u^1_{\theta} }{\omega_p^2} \ ,
\\ \nonumber
 \varepsilon_{eff} &=& \varepsilon_r \left(\frac{\pi}{2},\phi \right) +
 U \ ,
\\ \nonumber
 -U' &=&
  u^0_{\theta}
  - \frac{f^1_{\theta} u^0_{\phi} + u^1_{\theta} f^0_{\phi} +
  u^1_{\theta} u^0_{\phi}}{\omega_p} +
  \frac{u^1_{\theta} u^1_{\phi} u^0_{\phi}}{\omega_p^2} \ .
\end{eqnarray}
Equations (\ref{eq:effective_planar_equation}) and
(\ref{eq:effective_parameters}) constitute the first main result of
this paper.

In the presence of the current, $u \neq 0$, the right hand side of
(\ref{eq:effective_planar_equation}) contains additional ``effective
forces'' added to the $-{\varepsilon_r}'$ term. Since all functions
depend on just one variable $\phi$, these forces can be always
represented as the derivatives of an additional energy $U$ according
to the definition (\ref{eq:effective_parameters}).

One of the forces, namely the $u^0_{\theta}$ term, requires a
special discussion. When $u \sim \sqrt{\varepsilon_r \omega_p}$ this
term becomes larger than ${\varepsilon_r}'$ on the right hand side
of Eq.~(\ref{eq:effective_planar_equation}), and the estimates for
$\dot\phi$ and $\ddot\phi$ made in
Sec.~\ref{sec:simple_E_zero_current} become invalid. As it was
discussed above, this means that in a general case with non-zero
$u^0_{\theta}$ the effective equations
(\ref{eq:effective_planar_equation}, \ref{eq:effective_parameters})
can be only used for currents $u \lesssim \varepsilon_r$. However,
if $u^0_{\theta}$ is identically equal to zero, while the other spin
torque terms in $U$ and $\alpha_{eff}$ remain non-zero, one can
apply Eqs.~(\ref{eq:effective_planar_equation},
\ref{eq:effective_parameters}) for currents up to $u \sim
\sqrt{\varepsilon_r \omega_p}$.

Corrections to the friction coefficient are explicitly dependent on
the current magnitude. In the presence of spin torque the sign of
the friction coefficient may change \cite{bazaliy:2007:APL,
bazaliy:2007:PRB, boj:2008, bazaliy:2009:SPIE} , reflecting the
possible influx of the energy from the current source into the
system.

Below we investigate the application of the effective planar
equation to the device with a ``tilted polarizer'' geometry.

\section{Tilted polarizer device}\label{sec:tilted_polarizer_device}

In tilted polarizer devices vector $\bf s$ is pointed at an angle
$\theta_s$ to the axis $z$ and constitutes an angle $\pi/2 -
\theta_s$ with the easy plane. We will assume that $\bf s$ is in the
$(x,z)$ plane, i.e., $\phi_s = 0$.

To calculate $u_{\theta}$ and $u_{\phi}$ from
Eq.~(\ref{eq:definitions_u_theta_phi}) one needs to know the
function $u({\bf n})$. In many cases \cite{slon96, slonczewski:2002}
it has a form
$$
u({\bf n}) = \frac{g_0 I}{1+g_1 ({\bf n}\cdot {\bf s})}
$$
We will consider the case of small $g_1 \ll 1$ and approximate
\begin{equation}\label{eq:g_definition}
u({\bf n}) = g_0 I (1 - g_1 ({\bf n}\cdot {\bf s}))
\end{equation}
Using the expressions in Appendix~\ref{appendix:vector_definitions},
we find
\begin{eqnarray*}
u_{\theta} &=& g_0 I  \left[
  1 - g_1 (\sin\theta_s\sin\theta\cos\phi +
  \cos\theta_s\cos\theta)
  \right]
\\
  && \times (\sin\theta_s\cos\theta\cos\phi -
  \cos\theta_s\sin\theta) \ ,
\\
u_{\phi} &=& - g_0 I \left[
  1 - g_1 (\sin\theta_s\sin\theta\cos\phi +
  \cos\theta_s\cos\theta)
  \right]
\\
&& \times  \sin\theta_s\sin\phi \ .
\end{eqnarray*}
Therefore
\begin{eqnarray}
 \nonumber
u^0_{\theta} &=& - g_0 I \left[
  1 - g_1 \sin\theta_s\cos\phi \right] \cos\theta_s \ ,
\\ \label{eq:u0_tilted_polarizer}
u^0_{\phi} &=& - g_0 I \left[
  1 - g_1 \sin\theta_s\cos\phi \right] \sin\theta_s \sin\phi .
\end{eqnarray}
and
\begin{eqnarray}
 \nonumber
u^1_{\theta} &=& -  g_0 I [ \sin\theta_s\cos\phi +
  g_1 (\sin^2\theta_s\cos^2\phi - \cos^2\theta_s)] \ ,
\\ \label{eq:u1_tilted_polarizer}
u^1_{\phi} &=&  - g_0 I g_1 \sin\theta_s\cos\theta_s\sin\phi \ .
\end{eqnarray}

\subsection{In-plane polarizer}

In the case of in-plane polarizer with $\theta_s = \pi/2$ further
simplifications happen:
\begin{eqnarray*}
u^0_{\theta} &=& 0
 \\
u^0_{\phi} &=& -(1-g_1\cos\phi)\sin\phi
 \\
u^1_{\theta} &=& -(1 - g_1\cos\phi)\cos\phi
 \\
u^1_{\phi} &=& 0
\end{eqnarray*}
As we see, the in-plane polarizer is one of the special cases with
$u^0_{\theta} = 0$, discussed at the end of
Sec.~\ref{sec:arbitrary_E_with_current}. Consequently, the effective
equation can be used up to the currents $u \sim g_0 I \sim
\sqrt{\varepsilon_r \omega_p}$. The coefficients
(\ref{eq:effective_parameters}) acquire the form
\begin{eqnarray}
 \nonumber
 m &=& \frac{1}{\omega_p} \ ,
\\
 \alpha_{eff} &=& \alpha +
 \frac{g_0 I  \big[2\cos\phi - g_1(3\cos^2\phi - 1)\big]}{\omega_p} \ ,
\\ \label{eq:parameters_in-plane_polarizer}
 -U' &=& - \frac{g_0 I (1-g_1\cos\phi)}{\omega_p} \times
\\ \nonumber
 & \times &  \left[
  g_0 I (1-g_1\cos\phi)\sin\phi\cos\phi \ -
  \right.
\\ \nonumber
 && \qquad \qquad \left. - (f^1_{\theta}\sin\phi + f^0_{\phi}\cos\phi)
  \right]  \ .
\end{eqnarray}
Importantly, the $\dot\phi^2$ term in
Eq.~(\ref{eq:effective_planar_equation}) vanishes identically.

In Refs.~\onlinecite{bazaliy:2007:APL, bazaliy:2007:PRB} the
in-plane polarizer was considered in the case of $g_1 = 0$ and
residual energy
\begin{equation}\label{eq:E_with_easy_axis_anisotropy_and_field}
\varepsilon_r = -\frac{\omega_a}{2}\sin^2\theta\cos^2\phi
 - h \sin\theta\cos\phi \ ,
\end{equation}
describing a device with small easy axis anisotropy $\omega_a \ll
\omega_p$ in an external magnetic field $h$, both pointed along the
$x$ axis. In this case one finds $f^1_{\theta} = f^0_{\phi} = 0$ and
expressions (\ref{eq:parameters_in-plane_polarizer}) reproduce the
results obtained in Refs.~\onlinecite{bazaliy:2007:APL} and
\onlinecite{bazaliy:2007:PRB}.

\subsection{General case of a tilted polarizer}

When the polarizer magnetization $\bf s$ points at an arbitrary
angle $\theta_s$ the term $u^0_{\theta}$ is nonzero and we have to
limit the current magnitudes to $g_0 I \lesssim \varepsilon_r$ to
maintain the validity of Eq.~(\ref{eq:effective_planar_equation}).
With smaller currents more terms can be discarded from the effective
equation without changing its accuracy. Parameter expressions
(\ref{eq:effective_parameters}) reduce to
\begin{eqnarray}
 \nonumber
 m &=& \frac{1}{\omega_p}  \ ,
\\ \label{eq:effective_parameters_u_order_E}
 \alpha_{eff} &=& \alpha -
 \frac{ (u^0_{\phi})' + u^1_{\theta} }{\omega_p} \ ,
\\ \nonumber
 -U' &=&  u^0_{\theta}  \ .
\end{eqnarray}
Moreover, for $g_0 I \lesssim \varepsilon_r$ the nonlinear term with
$\dot\phi^2$ becomes small enough to be dropped from
Eq.~(\ref{eq:effective_planar_equation}).

The effective planar equation now reads
\begin{equation} \label{eq:effective_planar_equation_tilted_polarizer}
\frac{\ddot\phi}{\omega_p} + \alpha_{eff}\dot\phi
 = -\frac{\partial\varepsilon_{eff}}{\partial\phi} \ ,
\end{equation}
where the effective damping and the effective energy can be
expressed through the polarizer tilting angle $\theta_s$ using
Eqs.~(\ref{eq:u0_tilted_polarizer}), (\ref{eq:u1_tilted_polarizer}),
and (\ref{eq:effective_parameters_u_order_E})
\begin{eqnarray}
 \nonumber
 \alpha_{eff} &=& \alpha + \frac{g_0 I}{\omega_p} (2 \sin\theta_s\cos\phi
     - g_1[3 \sin^2\theta_s \cos^2\phi - 1]) \ ,
\\ \nonumber
 \varepsilon_{eff} &=& \varepsilon_{r}(\frac{\pi}{2},\phi) +
   g_0 I (\cos\theta_s \cdot \phi
   - g_1 \sin\theta_s\cos\theta_s\sin\phi) \ .
 \\ \label{eq:effective_parameters_u_order_E_through_thetaS}
 &&
\end{eqnarray}

\subsection{Switching diagram of the tilted polarizer device}

Let us now discuss the consequences of the modification of $\alpha
\to \alpha_{eff}$ and $\varepsilon_r \to \varepsilon_{eff}$ in the
presence of spin torque. The advantage of the effective planar
approximation is the possibility of using the analogy with the
particle motion which enables one to use mechanical intuition to
qualitatively predict the behavior of the solutions of
Eq.~(\ref{eq:effective_planar_equation_tilted_polarizer}), and thus
understand the dynamics of the spin-transfer device.

We will assume the standard nanopillar device described by the
residual energy (\ref{eq:E_with_easy_axis_anisotropy_and_field}). In
the special case of an in-plane polarizer this problem was discussed
in our earlier publications.\cite{bazaliy:2007:APL,
bazaliy:2007:PRB, bazaliy:2009:SPIE} Modifications of the effective
damping transform the particle motion qualitatively when
$\alpha_{eff}(\phi)$ changes sign. Modifications of
$\varepsilon_{eff}(\phi)$ become qualitatively important when
equilibrium points appear or disappear as the energy profile is
deformed.

\begin{figure}[t]
\includegraphics[scale=0.4]{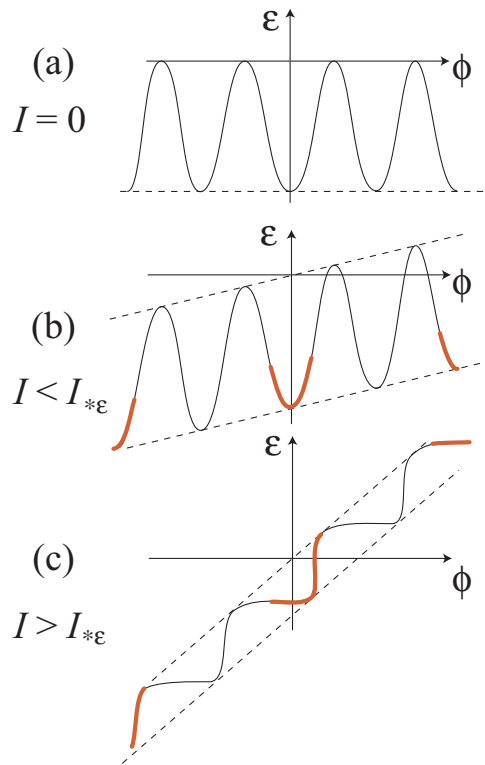}
\caption{Effective energy profile in the case of a tilted
polarizer.}
 \label{fig:energy_profile}
\end{figure}

Our goal here is to generalize the results of
Refs.~\onlinecite{bazaliy:2007:APL, bazaliy:2007:PRB,
bazaliy:2009:SPIE} to the case of a non-zero polarizer tilt and show
that the effective planar approach allows one to understand the
qualitative picture of the motion without doing the detailed
calculations. The case of a tilted polarizer was considered in
recent publications \cite{Zhou:2009, He:2010} using conventional
methods. The switching diagrams were calculated assuming that $\bf
s$ is tilted straight up from the easy axis direction. This
corresponds to our assumption of $\phi_s = 0$. It was also assumed
that there is no magnetic field, $h = 0$. We adopt the assumptions
of Refs.~\onlinecite{Zhou:2009} and \onlinecite{He:2010} to
illustrate the application of the formalism developed in this paper.

In the case of a general tilt angle $\theta_s$ the terms containing
a small factor $g_1$ give negligible corrections in the expressions
(\ref{eq:effective_parameters_u_order_E_through_thetaS}) for
$\alpha_{eff}$ and $\varepsilon_{eff}$. They can only become
important when $\theta_s$ approaches zero or $\pi/2$ and the main
terms vanish. We will assume that $\bf s$ is not to close to either
the in-plane or the perpendicular directions and an inequality
$\cos\theta_s, \sin\theta_s \gg g_1$ holds. Then we can use the
simplified form of
Eqs.~(\ref{eq:effective_parameters_u_order_E_through_thetaS})
\begin{eqnarray}
 \nonumber
 \alpha_{eff} &=& \alpha + \frac{2 g_0 I \sin\theta_s}{\omega_p} \cos\phi \ ,
\\ \label{eq:effective_parameters_u_order_E_through_thetaS_zero_g1}
 \varepsilon_{eff} &=& \varepsilon_{r}(\frac{\pi}{2},\phi) +
   g_0 I \cos\theta_s \cdot \phi \ .
\end{eqnarray}
Finding the critical currents in the narrow bands of angles
$\theta_s \approx 0$ or $\approx \pi/2$ where the $g_1$ terms are
important would require a more careful consideration.

\begin{figure}[t]
\includegraphics[scale=0.38]{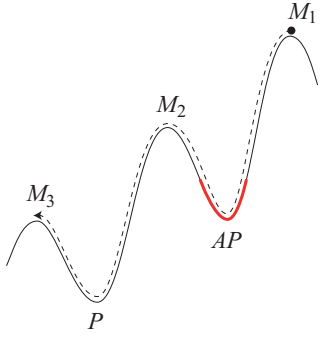}
\caption{Effective particle performing a 360$^\circ$ motion,
starting from the energy maximum $M_1$ and reaching an equivalent
maximum $M_3$. The red (gray) interval denotes the region of
$\alpha_{eff} < 0$.}
 \label{fig:OPP_process}
\end{figure}

The profile of the energy $\varepsilon_{eff}(\phi)$ is shown in
Fig.~\ref{fig:energy_profile}. The spin torque produces the tilted
washboard potential in which the effective particle moves. The
washboard tilt reflects the fact that spin torques can change the
magnetic energy of the system by using work performed by the current
source. The energy minima are shifted from their zero current
positions at $\phi = 0$ (parallel, or $P$ state) and $\phi = \pi$,
(antiparallel, or $AP$ state) to the new positions $\phi_0(I)$ that
are given by the equation
$$
\sin 2\phi_0 = \frac{2 g_0 I \cos\theta_s}{\omega_a} \ .
$$
One can check that the energy minima located at the angles
$\phi_{0min} + \pi n$ are still separated by the energy maxima
located at at $\phi_{0max} = \pi/2 - \phi_{0min} + \pi n$
(Fig.~\ref{fig:energy_profile}b). To be concise will still call the
$\phi_{0min}$ minimum the $P$-point, and the $\phi_{0min} + \pi$
minimum the $AP$ point.

As the current grows, the washboard tilts more and more, until the
extrema or $\varepsilon_{eff}(\phi)$ disappear altogether
(Fig.~\ref{fig:energy_profile}c). A short calculations shows that
this occurs at a critical current
\begin{equation}\label{eq:I_star_epsilon}
I_{*\varepsilon} = \frac{\omega_a}{2 g_0 \cos\theta_s} \ .
\end{equation}
For $I > I_{*\varepsilon}$ the effective particle slides down the
slope of the potential energy profile regardless of the sign and
magnitude of $\alpha_{eff}$. This corresponds to $\bf n$ performing
full 360$^\circ$ rotations in the azimuthal angle $\phi$. In the
spin transfer literature such a regime is called the out-of-plane
(OPP) precession.

Importantly, the OPP precession can exist even at $I <
I_{*\varepsilon}$. When the particle moves down the washboard, the
drop of its potential energy during one spatial period may be large
enough to overcome the friction energy loss
(Fig.~\ref{fig:OPP_process}). Therefore there must be a second
critical current $I_{OPP} < I_{*\varepsilon}$, such that the OPP
precession happens for $I > I_{OPP}$. In the interval $I_{OPP} < I <
I_{*\varepsilon}$ the stationary equilibrium of the particle at the
energy minimum coexists with the state of OPP precession. The
functional form of $I_{OPP}(\theta_s)$ depends on the energy profile
and the friction coefficient. Our goal here is not to find the
expression for it, but to see how far can we proceed in qualitative
understanding of the device dynamics without doing the actual
calculations.

\begin{figure}[t]
\includegraphics[scale=0.4]{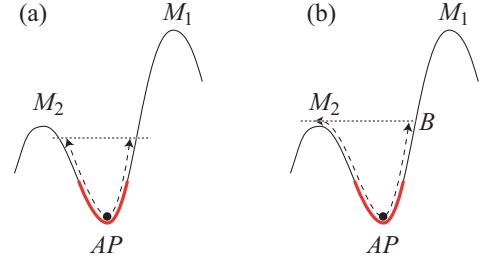}
\caption{(a) Effective particle performing small oscillations (IPP
precession) near the minimum corresponding to the AP state that lies
within he red (gray) interval of $\alpha_{eff} < 0$. (b) With
growing current the amplitude of the oscillations increases and the
particle reaches the $M_2$ maximum point. Above this threshold the
IPP precession is unstable.}
 \label{fig:IPP_process}
\end{figure}

The effective damping $\alpha_{eff}$ changes its sign in the
vicinity of the angles $\phi_{+n} = 2\pi (n + 1/2)$ for $I > 0$ and
$\phi_{-n} = 2\pi n$ for $I < 0$ when the current magnitude exceeds
the threshold
\begin{equation}\label{eq:I_star_alpha}
I_{*\alpha} = \frac{\alpha \omega_p}{2 g_0 \sin\theta_s} \ .
\end{equation}
The regions of $\alpha_{eff} < 0$ are shown in
Figs.~\ref{fig:energy_profile},\ref{fig:OPP_process} by red (gray)
thicker line. On the intervals of negative friction the system
acquires energy instead of loosing it. This makes it easier for the
particle to achieve the state of the OPP precession. The actual
calculation of the $I_{OPP}$ threshold must take into account the
energy gain due to the tilt of the potential and the presence of
negative friction intervals. Both features mathematically represent
the ability of the spin torques to transfer energy from the current
source to the system.

Another important effect of negative $\alpha_{eff}$ is the local
destabilization of the energy minima. If the minimum point
$\phi_0(I)$ lies within the interval of $\alpha_{eff} < 0$, it
becomes unstable and small oscillations around it are developed.
Fig.~\ref{fig:IPP_process}a shows such oscillations near the AP
minimum which, according to
Eq.~(\ref{eq:effective_parameters_u_order_E_through_thetaS_zero_g1}),
is destabilized for $I > 0$. These oscillations correspond to the
precession of $\bf n$ around the equilibrium point and are called an
in-plane (IPP) precession in the spin transfer literature.

The amplitude of the oscillations is determined by the balance of
energy influx and dissipation on the intervals of negative and
positive friction.\cite{bazaliy:2007:APL, bazaliy:2007:PRB,
bazaliy:2009:SPIE} As the current is increased, the amplitude grows
and eventually becomes so large that the particle reaches the crest
of the potential (Fig.~\ref{fig:IPP_process}b) and falls down into
the neighboring valley. This process leads to the destruction of the
IPP state. The latter therefore exists between the two threshold
currents. For the $AP$ equilibrium these are the $I_{AP}$ threshold,
where the $AP$ point becomes unstable, and the $I_{IPP}$ threshold,
where the precession amplitude becomes too large to be contained in
the $AP$ valley.

The critical current of minimum point destabilization is determined
from the equation $\alpha_{eff}(\phi_0(I)) = 0$ which can be
rewritten as
\begin{eqnarray}
 \nonumber
&& I_{AP} \cos[\phi_0(I_{AP})] = I_{*\alpha} \ ,
 \\ \label{eq:I_P_AP_equation}
&& I_{P} \cos[\phi_0(I_{P})] = - I_{*\alpha} \ ,
\end{eqnarray}
for $P$ and $AP$ minima. The equations show that the threshold
currents satisfy $|I| > I_{*\alpha}$ in both cases. This result can
be naturally understood as follows. The friction first becomes
negative at the $\phi = 0$ or $\phi = \pi$ points at $I = \pm
I_{*\alpha}$. But the $P$ and $AP$ minima are shifted from the $0,
\pi$ points to the $\phi_0(I)$ points. In order to destabilize them
the negative friction interval has to grow large enough to cover the
actual minima positions.

The critical current $I_{IPP}$ depends on the shape of the potential
in the entire interval traveled by the particle in
Fig.~\ref{fig:IPP_process}b. Its actual calculation is not the goal
of our qualitative approach. We can make two general statements
about $I_{IPP}$. First, the destabilization of the precession state
certainly happens at a current that is larger than the one required
for the destabilization of the corresponding energy minimum. For
example, $I_{IPP} > I_{AP}$.

\begin{figure}[t]
\includegraphics[scale=0.45]{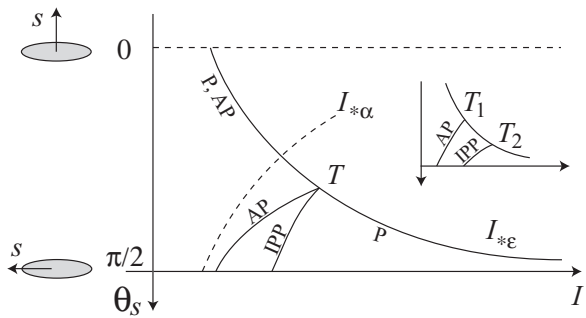}
\caption{Critical lines of the $P/AP$ equilibrium states and the IPP
precession state on the $(I,\theta_s)$ switching diagrams. Each line
is marked by the name of the state which gets destabilized on it.
The name is put on the side of the line where the state is stable.
Inset: an impossible arrangement of the $AP$ and $IPP$ lines.}
 \label{fig:switching_diagram_preliminary}
\end{figure}

To set the stage for the second observation, we proceed to the
discussion of the switching diagram. In our case the experimental
parameters are the current $I$ and the tilting angle $\theta_s$.
Various critical currents are represented as lines on the
$(I,\theta_s)$ plane and divide it into domains with different sets
of stable states. The currents $I_{*\varepsilon}$, $I_{*\alpha}$,
$I_{AP}$, and $I_{APP}$ are sketched as functions of the angle
$\theta_s$ in Fig.~\ref{fig:switching_diagram_preliminary}. The
lines $I_{*\varepsilon}(\theta_s)$ and $I_{*\alpha}(\theta_s)$,
given by Eqs.~(\ref{eq:I_star_epsilon}) and (\ref{eq:I_star_alpha}),
intersect at a certain point. Due to $I_{AP} > I_{*\alpha}$ the
$I_{AP}(\theta_s)$ has to be located to the right of the
$I_{*\alpha}(\theta_s)$ line on the diagram. It intersects the
$I_{*\varepsilon}(\theta_s)$ line as well --- one can prove that
$I_{AP}(\theta_s)$ is a decreasing function of the tilt angle. The
$I_{IPP}(\theta_s) > I_{AP}(\theta_s)$ is also a decreasing function
and the corresponding line has to be located even further to the
right. Both $I_{AP}(\theta_s)$ and $I_{IPP}(\theta_s)$ cross the
$I_{*\varepsilon}(\theta_s)$ line, and one can prove that they do it
at the same point $T$. This is second general property of the
$I_{IPP}$ threshold.

\begin{figure}[b]
\includegraphics[scale=0.45]{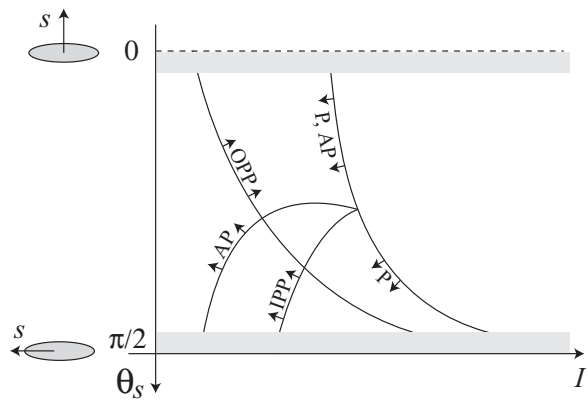}
\caption{Switching diagram of a spin-transfer device with a tilted
polarizer: a qualitative sketch. Each line is marked by the name of
the state which gets destabilized on it. The arrows next to the name
point to the side of the line where this state is stable. Gray
horizonthal strips denote the regions where the approximation
(\ref{eq:effective_parameters_u_order_E_through_thetaS_zero_g1}) may
fail and (\ref{eq:effective_parameters_u_order_E_through_thetaS})
should be used.}
 \label{fig:switching_diagram}
\end{figure}

The second property can be proven by considering a hypothetical
switching diagram shown in the inset in
Fig.~\ref{fig:switching_diagram_preliminary} where it is assumed
that that the $AP$ and $IPP$ lines cross
$I_{*\varepsilon}(\theta_s)$ at different points $T_1$ and $T_2$.
Consider the point $T_2$. Approaching it from within the domain of
existence of the $IPP$ precession one should observe a decreasing
amplitude of oscillations around the $AP$ minimum, because the size
of the valley around $AP$ shrinks to zero. At $T_2$ the amplitude of
sustained oscillations around $AP$ should be equal to zero. At the
same time the $AP$ point should be unstable because the current is
larger than the $I_{AP}$ threshold. But an instability of a minimum
with zero amplitude of oscillations is the definition of the
$I_{AP}$ current, so $T_2$ has to lie on the $AP$ line. This
argument proves that the assumption about the existence of two
different crossing points $T_{1,2}$ was inconsistent.

The relationships between $I_{AP}$ and $I_{IPP}$ discussed above
follow from the fact that both currents are determined by the energy
and friction near the same minimum. The $I_{OPP}$ current depends on
the details of $\varepsilon_{eff}$ and $\alpha_{eff}$ on the whole
$2\pi$ interval of the angle $\phi$ and thus no qualitative
relationships for it can be found, except for the already mentioned
$I_{OPP} < I_{*\varepsilon}(\theta_s)$. A qualitative sketch of the
full switching diagram is given in Fig.~\ref{fig:switching_diagram}.
The $I < 0$ part of the sketch can be obtained by reflecting it with
respect to the vertical axis: the diagram is symmetric with respect
to the $I \to - I$ transformation. This is a consequence of two
symmetries built into the energy and friction functions
~(\ref{eq:effective_parameters_u_order_E_through_thetaS_zero_g1}):
the $\pi$-periodicity of the energy
(\ref{eq:E_with_easy_axis_anisotropy_and_field}), and the
$\alpha_{eff}(\pi-\phi, - I) =  \alpha_{eff}(\phi, I)$ symmetry of
the effective friction. The latter depends on $\bf s$ being directed
in the symmetry plane and the fact that the $g_1$ terms were
dropped. If either of the two symmetries is violated, the $P$ and
$AP$ states would no longer be equivalent in all respects.

Notably, Fig.~\ref{fig:switching_diagram} reproduces all qualitative
features of the switching diagrams obtained in
Refs.~\onlinecite{Zhou:2009} and \onlinecite{He:2010} by
conventional methods. In addition, it brings important qualitative
understanding of the behavior of critical currents as a function of
system parameters and approximations used for the spin torque
efficiency factor $g({\bf n})$. For example, we see that the
threshold currents in the two most frequently considered limiting
cases of the perpendicular ($\theta_s = 0$) and in-plane ($\theta_s
= \pi/2$) polarizers will be most sensitive to the $g({\bf n})$
function used for the calculations.

Finally, we should mention that the results of the effective planar
approach are not limited to the qualitative conclusions discussed in
this paper. It allows one to calculate the critical currents, often
being the only one providing analytical expressions in the case of
precession states. In our case the $I_{*\varepsilon}$ threshold is
given by the expression~(\ref{eq:I_star_epsilon}). The $I_{P}$ and
$I_{AP}$ currents can be obtained from
Eqs.~(\ref{eq:I_P_AP_equation}). At the crossing point $\theta_{s*}$
of the $AP$, $IPP$ and $I_{*\varepsilon}$ lines both equations
should hold, which gives
$$
\tan\theta_{s*} = \frac{\sqrt{2}\alpha\omega_p}{\omega_a}
$$
In the limit of small friction $\alpha \ll \sqrt{\omega_a/\omega_p}$
considered here the critical currents $I_{OPP}(\theta_s)$ and
$I_{IPP}(\theta_s)$ related to the precession states may be found
using the method introduced in Ref.~\onlinecite{bazaliy:2007:PRB}.

\section{Conclusions}

We have given the detailed derivation of the effective planar
equation for spin-transfer devices with dominating easy plane
anisotropy and illustrated its application by performing a
qualitative study of a spin-transfer device with tilted polarizer.
Once the parameters of the effective equation are found, the
approach allows one to understand the dynamics qualitatively without
performing detailed calculations. This is especially important in
the case of precession cycles which are usually studied numerically.
The method also elucidates the role of approximations used to model
the spin torque and shows the limits of their applicability.

The obtained switching diagram demonstrates a competition between
the two types of switching. For small $\theta_s$ the destabilization
of the $AP$ minimum results from the merging and disappearance of
the minimum and maximum points of $\varepsilon_{eff}$. For
$\theta_s$ close to $\pi/2$ the destabilization happens locally,
changing the nature of the $AP$ equilibrium from stable to unstable.
This type of competition is not unique to the systems with strong
easy plane anisotropy --- it was shown in
Ref.~\onlinecite{sodemann:2011} that it may happen in any
spin-transfer device.

\begin{figure}[t]
    \resizebox{.22\textwidth}{!}{\includegraphics{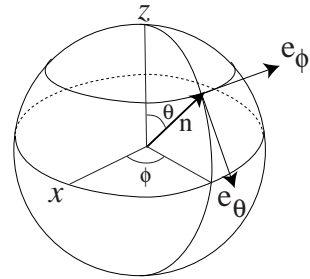}}
\caption{Definitions of the tangent vectors and polar angles.}
 \label{fig:vectors}
\end{figure}

\section{Acknowledgments}
This research was supported by the NSF grant DMR-0847159.

\appendix

\section{Vector definitions}
 \label{appendix:vector_definitions}

We use the standard definitions of polar coordinates and tangent
vectors (see Fig.~\ref{fig:vectors}):
\begin{eqnarray}
 \nonumber
{\bf n} &=& (\sin\theta\cos\phi, \sin\theta\sin\phi, \cos\theta)
 \\
 \label{eq:vector_components}
{\bf e}_{\theta} &=& (\cos\theta\cos\phi, \cos\theta\sin\phi,
-\sin\theta)
 \\
 \nonumber
{\bf e}_{\phi} &=& (-\sin\phi, \cos\phi, 0)
\end{eqnarray}
For the polarizer unit vector ${\bf s}$ with polar angles
$(\theta_s,\phi_s)$ the scalar product expressions are
\begin{eqnarray}
 \nonumber
 ({\bf s} \cdot {\bf e}_{\theta}) &=&
 \sin\theta_s\cos\theta\cos(\phi_s - \phi) -
 \cos\theta_s\sin\theta
 \\
  \label{eq:scalar_products}
 ({\bf s} \cdot {\bf e}_{\phi}) &=&
 \sin\theta_s \sin(\phi_s - \phi)
\end{eqnarray}

\end{document}